\documentclass[prd,floats,floatfix,showpacs,nofootinbib,dvipdfm]{revtex4}
\usepackage{graphicx}
\usepackage{dcolumn}
\usepackage{bm}
\usepackage{graphics}
\usepackage{slashed}
\usepackage{amssymb}
\usepackage{natbib}
\usepackage{amsmath}
\usepackage{url}
\usepackage{color}
\newcommand{\fracc}[2]{\frac{\textstyle{#1}}{\textstyle{#2}}}

\begin{document}

\title{A proposal for the origin of the anomalous magnetic moment}

\author{M. Novello}\email{novello@cbpf.br}
\author{E. Bittencourt}\email{eduhsb@cbpf.br}

\affiliation{Instituto de Cosmologia Relatividade Astrofisica ICRA -
CBPF\\ Rua Dr. Xavier Sigaud, 150, CEP 22290-180, Rio de Janeiro,
Brazil}

\date{\today}

\begin{abstract}
We investigate a new form of contribution for the anomalous magnetic
moment of all particles. This common origin is displayed in the
framework of a recent treatment of electrodynamics that is based in
the introduction of an electromagnetic metric which has no
gravitational character. This effective metric constitutes a
universal pure electromagnetic process perceived by all bodies,
charged or not charged. As a consequence it yields a complementary
explanation for the existence of anomalous magnetic moment for
charged particles and even for non-charged particles like neutrino.
\end{abstract}

\maketitle

\section{Introduction}

In the standard model of particle physics neutrinos possess the most
intriguing and interesting behavior. It is understood that neutrinos
have only gravitational and weak interaction. Although they do not
have a standard magnetic moment---since they have no charge---there
has been investigations beyond the standard model, concerning the
possibility that neutrinos could have an effective magnetic moment
(see details of this proposal in Refs.
\cite{sanda,lee,helayel,kau}). For any charged particle the
classical magnetic moment ($\mu$) is inversely proportional to the
mass $\mu=e\hbar/2m$, where $\hbar$ is the reduced Planck constant,
$e$ is the elementary charge and $m$ is the mass of the particle. In
the case of electrons, for example, $\mu\equiv\mu_B = e\hbar/2m_e$
is the Bohr magneton. However, in the case of neutrinos $\nu$, it
has been suggested that $\mu_{\nu}$ depends linearly on the neutrino
mass, due to some restrictions on the extension of the standard
model. As a consequence, this type of coupling between the spinor
field and the electromagnetic field presents some difficulties
concerning the variational principle and the re-normalization
process.

In this paper we argue that the magnetic moment of any particle is
composed of two very distinct parts: the standard one that depends
on the charge of the particle---from dimensional analysis, it is
inversely proportional to the corresponding mass---and another part
(we call the geometrical magnetic moment) that depends linearly on
the mass. This second case, which is many orders of magnitude lower
than the standard one, is common for all particles and does not
depend on their charge. In other words, we shall see that part of
the anomalous magnetic moment of charged particles (we deal here in
the leptonic sector) is a consequence of such geometrical magnetic
moment and for neutral particles, like neutrinos, our proposal
provides a nonzero magnetic moment. We shall compare the effects of
both terms from experimental data.

Our proposal is based on Ref. \cite{novello et al}, in which the
equivalence between the electromagnetic Born-Infeld theory in a
specific curved geometry and Maxwell's theory in the Minkowski
space-time was demonstrated. In other words Maxwell's linear
electrodynamics in flat space-time may be deformed into a
Born-Infeld theory in a curved space-time. This means that any
solution of the former is also a solution of the latter, which is
given in terms of a prescribed map. This is possible because the
associated curved space-time depends only on the electromagnetic
fields. Indeed, due to the algebraic structure of the
electromagnetic two-form $F_{\mu\nu}$ and its dual, there exist a
kind of closure relation that allows the existence of this mapping,
thus generating a dynamical bridge between these two paradigmatic
theories.

We will show that the results presented in that paper can become
more than a simple mathematical analogy. Indeed the application of
the dynamical bridge in specific situations suggests a new manner to
understand the origin of the anomalous magnetic moment, which
depends linearly on the mass, as a direct consequence of the
electromagnetic geometry.

\section{The Electromagnetic Dynamical Bridge}

Recently, it was shown that the Born-Infeld dynamics in a curved
space-time endowed with an associated metric $ \hat e_{\mu\nu}$ is
dynamically equivalent to the linear Maxwell electrodynamics
described in a flat Minkowski background \cite{novello et al}.
Namely, there is a map, which we call dynamical bridge (DB),
relating these two dynamics in such a way that they are distinct
representations of one and the same physics. It means that
\textit{any} solution of the former will also be a solution of the
latter, which is given in terms of a preestablished map. This task
is highly nontrivial and  becomes possible only if the space-time
metric depends explicitly on the electromagnetic field. Due to the
algebraic structure of the electromagnetic tensor $F_{\mu\nu}$ and
its dual ${}^*F_{\mu\nu}$ (see details in Appendix \ref{apb}), there
are some closure relations that allow the existence of this map,
generating a connection between those two paradigmatic theories. The
apparent disadvantage is to leave aside the simple Minkowski
background $\eta_{\mu\nu}$ and to go to a specific curved space-time
$\hat{e}_{\mu\nu}$, which is constructed solely in terms of the
background metric and the electromagnetic fields. In principle, one
could suspect that this map is useless because it links a very
simple electromagnetic theory in flat space to a nonlinear theory in
a curved space-time. Notwithstanding, we will show that, according
to the particular situation physical principles will guide us in the
choice of the more appropriate representation to describe the system
under consideration.

The motivation to deal with the Born-Infeld theory in curved space
is similar to the one adopted by Born and Infeld in their seminal
paper \cite{born}, whose the main idea consists in the production of
a scenario where the classical electromagnetic field is finite; in
practical means, the theory has a ``cut off" on a critical value for
the electromagnetic field.

Besides, the introduction of a curved geometry given entirely in
terms of the electromagnetic field requires, from the physical point
of view, the presence of this characteristic (the critical field) in
order to guarantee the regularity of the curvature of the
space-time. In this vein, let us briefly review the DB method
applied to this scenario making a self-consistent the exhibition of
our proposal.

We start by considering the Born-Infeld Lagrangian for the
electromagnetic field $F_{\mu\nu}$, defined in a curved geometry
$\hat e_{\mu\nu}$, as follows

\begin{equation}
\label{l_bi}
L=\beta^2\left(1-\sqrt{\hat U}\right),
\end{equation}
where $\beta$ is a free parameter interpreted as the critical
electromagnetic field and
$$\hat U\equiv1+\fracc{\hat F}{2\beta^2}-\fracc{\hat G^2}{16\beta^4}.$$
The dynamical equation for the electromagnetic field in the
Born-Infeld theory is
\begin{equation}
\label{din_bi}
\partial_{\nu}\left[ \frac{\sqrt{- \, \hat e }}{\hat U } \left(  \hat F^{\mu\nu} - \frac{1}{4 \, \beta^{2}} \, \hat G \, {}^*\hat F^{\mu\nu}\right)  \right] = 0,
\end{equation}
where $\hat{e}$ is the determinant of the metric $\hat e_{\mu\nu}$.
Let us define a particular curved space-time endowed with the
electromagnetic metric defined as
\begin{equation}
\hat e^{\mu\nu} \equiv a \, \eta^{\mu\nu} + b \, \Phi^{\mu\nu},
\end{equation}
and
$$ \Phi_{\mu\nu} \equiv F_{\mu\alpha} \, F^{\alpha}{}_{\nu}.$$

The coefficients $a$ and $b$ of the electromagnetic metric tensor
should be functions of the invariants $F \equiv F_{\mu\nu} \,
F^{\mu\nu}$ and $G \equiv  F^{\mu\nu} \, {}^*F_{\mu\nu}$ construct
with the tensor $F_{\mu\nu}$ and its dual. The algebraic relation
between $\Phi_{\mu\nu}$ and $F_{\mu\nu}$ is unique. Any other
expression for $\Phi_{\mu\nu}$ constructed in terms of $F_{\mu\nu}$
and its dual is a combination of $\eta_{\mu\nu}$ and
$F_{\mu\alpha}F^{\alpha}{}_{\nu}$, which can then be incorporated
into the coefficients $a$ and $b$. The indexes of all quantities
with the symbol ($\,\hat\,\,$) are raised and lowered with the
metric $ \hat{e}_{\mu\nu}$. In particular, the invariants $\hat F$
and $ \hat G $ are constructed with this metric\footnote{See
Appendix \ref{apb} to find the definition of the quantities that
appear in this section, like $\hat U$, $\hat F$, $\hat G$ etc.}.

To show this equivalence, it is worth to evaluate the tensor $\hat
F^{\mu\nu}$ and its dual, defined in the electromagnetic metric
$\hat e_{\mu\nu}$ as functions of the electromagnetic field $
F^{\mu\nu}$ and its dual constructed in the Minkowski geometry.

The definition of $\hat F^{\mu\nu}$ and the choice of the metric
$\hat e_{\mu\nu}$ implies that these fields are related through the
expression

\begin{equation}
\label{transf_campos}
\left(\begin{array}{c}
\fracc{\hat F^{\mu\nu}}{a^2}\\[1.5ex]
\fracc{^*\hat F^{\mu\nu}}{a^2}
\end{array}\right)=
\left(\begin{array}{cc}
p-\epsilon F q&-\fracc{\epsilon Gq}{2}\\[2ex]
-\fracc{\epsilon Gq}{2}&p
\end{array}\right)
\left(\begin{array}{c}
F^{\mu\nu}\\[2ex]
^*F^{\mu\nu}
\end{array}\right),
\end{equation}
where $\epsilon\equiv b/a$ and
$$p=1+\fracc{\epsilon^2G^2}{16},\hspace{1cm}q=1-\fracc{\epsilon F}{4}.$$

We can interpret Eq.\ (\ref{transf_campos}) as nothing but a map
from $ (F_{\mu\nu},  F_{\mu\nu}^{*}) $ into $ (\hat{F}_{\mu\nu},
\hat{F}_{\mu\nu}^{*}). $ Substituting this equation in Eq.\
(\ref{din_bi}) and making the requirement that Born-Infeld dynamics
in the $ \hat{e}_{\mu\nu} $ corresponds to the Maxwell dynamics in
flat Minkowski space-time, we obtain the following equations for the
coefficient of the electromagnetic metric (see \cite{novello et al}
for more details)

\begin{equation}
\label{eqs_din_eq}
\left\{\begin{array}{lcl}
p-\epsilon Fq+\fracc{\hat G}{2\beta^2}\epsilon q&=&-\fracc{Q}{4},\\[2ex]
-\epsilon Gq+\fracc{\hat G p}{2\beta^2}&=&0,
\end{array}\right.
\end{equation}
where
$$Q=1-\fracc{\epsilon F}{2}-\fracc{\epsilon^2G^2}{16}.$$
Solving these equations for $a$ and $\epsilon$, we get the final
form of metric $\hat e_{\mu\nu}$ which maps the solutions of the
Born-Infeld theory in the $ \hat{e}_{\mu\nu} $ curved space into
solutions of the Maxwell theory in Minkowski flat space. Although
this is a very general result we will consider here only the case in
which the topological invariant $ G $ vanishes. We postpone the
analysis of the general case for another place. This regime
corresponds to an electric field orthogonal to the magnetic field in
the Maxwell theory (which includes plane waves). From this
assumption, the second equation of\ (\ref{eqs_din_eq}) is
identically satisfied. Therefore, one of the metric coefficients
remains undetermined and, for simplicity, we fix $a=1$. The other
one is determined by equation (\ref{eqs_din_eq}) and is given by the
roots of a third order polynomial. There are 3 possibilities for the
values of $\epsilon:$

\begin{equation}
\epsilon = \left\{\epsilon_0,\epsilon_+,\epsilon_-\right\}=
 \left\{\fracc{2}{F},
 \fracc{2}{F}\left(1\pm\fracc{1}{\sqrt{1-F/2\beta^2}}\right)\right\}.
\end{equation}
We note that $\epsilon_0$ can be neglected, once in this case the
determinant of the matrix relating the fields in Eq.\
(\ref{transf_campos}) vanishes.

Consider the regime in which the electromagnetic field is very low
in comparison to the critical field, i.e., $F\ll\beta^{2}$. In this
case, if we perform a power series expansion in terms of
$F/2\beta^2$ for the remaining allowed solutions, we see that the
series does not converge for $\epsilon_+$. Therefore, only
$\epsilon_-$ is an acceptable solution in this regime and the
electromagnetic metric, in first order of $F/2\beta^{2}$, reduces to
the form
\begin{equation}
\hat e_{\mu\nu} \approx \eta_{\mu\nu} + \frac{1}{2 \, \beta^{2}} \, \Phi_{\mu\nu}.
\label{1721}
\end{equation}
From now on, we shall examine only this regime for the metric.
Remark that it is possible to interpret the presence of the
electromagnetic metric as a small deviation from the flat space-time
generated by the electromagnetic field. From the dynamical bridge
between the equivalent representations it follows that we can
describe pure electromagnetic phenomenon either in terms of Maxwell
theory in flat Minkowski space-time or as Born-Infeld dynamics in
the electromagnetic curved geometry.

Henceforth we denote Maxwell-Minkowski or MM-representation the case
where one chooses to describe electromagnetic processes in the
linear Maxwell theory in the flat Minkowski space-time. On the other
hand, when one applies the dynamical bridge approach and describes
the same processes in the nonlinear Born-Infeld theory in a curved
space-time driven by the metric $\hat{e}_{\mu\nu}$, it will be
denoted electromagnetic metric or for short, the $\hat E-$
representation. Let us emphasize that these representations describe
one and the same dynamics.

\section{The electromagnetic metric}
The presence of a curved structure of the space-time in the realm of
electromagnetic fields can be an important theoretical instrument of
analysis only if one introduces a prescription of how matter
perceives this geometry. Then, the question is: how matter interacts
with the electromagnetic metric? This question appears immediately
as long as one wants to explore this metric formulation in the
presence of matter.

At this point, we are led to propose the natural idea that
\textbf{all kind of particles, charged or not charged, interacts
with the metric $ \hat e_{\mu\nu}$ in a universal and unique way.}
The simplest manner to realize this hypothesis is by using the
minimal coupling principle that we borrow from gravitational
processes as they are described in general relativity.  Although
this hypothesis may appear at first glance very unusual and even an
heresy, we shall see that it contains a well-posed program with
observational consequences that could prove or disprove it. In order
to proceed with this idea we examine the case of spinor fields in
particular the leptonic sector of the standard model, to investigate
what new effects one can extract from this geometrical scenario.

The extended dynamical bridge concerns the behavior of all kind of
matter. In the $\hat E$-representation the assumption of complete
democracy---that is, the idea that any kind of matter,
charged-or-not, lives in the geometry $\hat e_{\mu\nu}$---implies an
equivalent effect in the MM-representation. In particular, the
equation of motion of the neutrino in the electromagnetic geometry,
provokes the need to assume a direct interaction between the
neutrino and the electromagnetic field in the MM-representation.
Indeed, one is led
to accept that in this representation there should exist an extra term
in the Lagrangian which is the analogue of the minimal coupling
principle between electromagnetic field and the neutrino embedded in
the metric $\hat e_{\mu\nu}$. We shall see that, for instance, the role of the
parameter $\beta$ that appears explicitly in the $\hat
E$-representation, becomes the ratio between the magnetic moment of
any particle due to its interaction with the electromagnetic metric
and its corresponding mass.

\section{Minimal coupling principle}

Let us describe the electromagnetic effects on the spinor field
equation in the $\hat E$-representation. We explore the
 equivalence displayed by the dynamical bridge and deal with an
extended version of the way matter interacts with the
electromagnetic field. It is a universal and well-accepted belief
that only charged matter is able to couple directly with the
electromagnetic field. Nevertheless, from what we have shown above,
the analysis in the $\hat{E-}$ representation allows a new
possibility of interaction.

The equivalence between the Born-Infeld theory in the $\hat
E$-representation and the Maxwell dynamics in flat space was shown
only in the case of free fields. We extend this equivalence by
assuming that matter couples universally to the electromagnetic
field in the $\hat e_{\mu\nu}$ framework. The question then is: what
are the consequences of this hypothetical universality if the
electromagnetic process is described in the MM-representation?

Following our approach, let us investigate the consequences of the
coupling in the $\hat{E}$-representation. We start by noticing that
there are two possible forms of interaction of the matter with the
electromagnetic field, that is
\begin{itemize}
\item{through the vector potential $A_{\mu}$;}
\item{through the metric tensor $\hat e_{\mu\nu}$.}
\end{itemize}

Once only the geometrical form of interaction was not considered
before, we will concentrate our analysis here on this second way.
Let us point out that the existence of the electromagnetic geometry
makes sense and may have further consequences only if this geometry
is perceived for all kind of matter. We are then led to state the
following assertion, that may be felt at first glance as an heresy:
if in a given domain there exists an electromagnetic field, then
every single body perceives itself as embedded in the
electromagnetic geometry $\hat e_{\mu\nu}$. This means that a test
particle can couple to the electromagnetic field without having
electric charge. In other words, if a particle has an electric
charge it couples with the electromagnetic field through the
standard channel via the potential $A_{\mu}.$ This happens only for
the class of bodies that are denominated \emph{charged} particles.
On the other hand, all particles (charged or not) interacts with the
electromagnetic geometry in an unique and the same way, according to
the minimal coupling principle stated above. What are the main
consequences of this apparently strange hypothesis? That is the
purpose of the next section.

\section{The $\hat{E}-$metrical origin of the neutrino magnetic moment}

In the case of neutrinos, which are uncharged, there is only the
geometrical way to couple with the electromagnetic field. We will
apply the minimal coupling principle in the $\hat E$-representation
and analyze the consequences of this in the standard
Maxwell-Minkowski representation.

We start by defining the associated Dirac matrices $ \hat
\gamma^\alpha $ by the relation
\begin{equation}
\{\hat \gamma^{\mu},\hat \gamma^{\nu}\}= 2 \, \hat e^{\mu\nu} \, \textbf{1},
\end{equation}
where $ \textbf{1} $ is the identity matrix of the associated
Clifford algebra and the curly brackets means anti-commutation.
Analogously, in the MM-representation, we have
\begin{equation}
\{\gamma^{\mu},\gamma^{\nu}\}= 2 \, \eta^{\mu\nu} \, \textbf{1}.
\end{equation}

It then follows that we can set\footnote{The most general expression
for the $\hat\gamma^{\mu}$ in terms of the Dirac matrices
$\gamma^{\mu}$ and the electromagnetic field $F_{\mu\nu}$ is
$$\hat\gamma^{\mu} = \gamma^{\mu} + [\lambda\sqrt{\epsilon} F^{\mu}{}_{\alpha} + q \epsilon\Phi^{\mu}{}_{\alpha}] \gamma^{\alpha},$$
where $\lambda$ and $q$ are constants satisfying $2q=\lambda^2+1$
and $\epsilon=-1/2\beta^2$. To simplify our exposition we set here
$\lambda=0$.}

\begin{equation}
\label{gamma_hat}
\hat\gamma^{\mu}=\gamma^{\mu}- \frac{1}{4\, \beta^{2}} \Phi^{\mu}{}_{\alpha}\gamma^{\alpha}.
\end{equation}
Therefore, the dynamical equation for the spinor field, in the metric $\hat e_{\mu\nu}$, becomes
\begin{equation}
i \hbar c\, \hat \gamma^{\mu} \hat \nabla_{\mu} \, \Psi -  m c^{2}\, \Psi = 0. \label{746}
\end{equation}
The covariant derivative $\hat \nabla_{\mu} \equiv
\partial_{\mu}-\hat \Gamma_{\mu}$ is given in terms of the
Christoffel symbols constructed with the electromagnetic metric and
the internal connection is provided by the Fock-Ivanenko
coefficients obtained from the generalized Riemannian condition
\begin{equation}
\hat \nabla_{\mu} \, \hat \gamma^{\nu} = [ \hat V_{\mu},  \hat \gamma^{\nu} ], \label{1235}
\end{equation}
where $\hat V_{\mu}$ is an arbitrary element of the associated
Clifford algebra \cite{novello_73}. Note that Eq.\ (\ref{1235}) is
the most general condition upon the Dirac matrices that allows a
Riemannian geometry satisfying the condition that the covariant
derivative of the metric vanishes, i.e., $ \hat \nabla_{\alpha} \,
\hat e^{\mu\nu} = 0$. In the Appendix \ref{apa}, one can see that,
even under this general condition, the conservation law is still
satisfied in both representations.

The presence of $V_{\mu}$ suggests that the interaction between the
electromagnetic field  and $\Psi$ occurs in the internal space. In
the absence of any kind of matter, we are free to assume that the
commutator on the right-hand side of Eq.\ (\ref{1235}) vanishes.
However, when matter (of any kind) exists, $\hat V_{\mu}$ depends
simultaneously on the electromagnetic field and on the properties of
the matter field and, in the case of spinors, we set
\begin{equation}
\label{choice_v} \hat V_{\mu} = i\frac{m \, c}{\hbar \,\beta} \,
F_{\mu\nu} \, \hat \gamma^{\nu}\gamma_5,
\end{equation}
where $m$ is the neutrino mass in the minimal extended version of
the standard model. In order to guarantee that we are dealing only
with uncharged particles, we assume that the Fock-Ivanenko
connection does not have any term proportional to the identity ${\bf
1},$ according to \cite{novello_73}. It should be remarked that even
if we consider massless neutrinos (which is indeed the standard
model assumption), we can rewrite $\hat V_{\mu}$ is terms of the
neutrino energy and then the coupling to the electromagnetic field
is still present. This replacement of $m$ by $E$ suggests that such
interaction should be examined in order to investigate the existence
of the right-handed neutrinos \cite{shapos,geng}.

Using Eqs.\ (\ref{1721}) and\ (\ref{choice_v}) into the equation of motion (\ref{746}), we get

\begin{equation}
i \, \hbar c \, \gamma^{\mu} \partial_{\mu} \, \Psi  + \frac{\,m c^{2}}{\beta} \,
F_{\mu\nu} \, \sigma^{\mu\nu}\gamma_5 \, \Psi - m \, c^{2} \, \Psi = 0,
\label{1730}
\end{equation}
which, according to the regime under consideration, corresponds to
the first order approximation in $F/2\beta^2$
where $
\sigma^{\mu\nu} = (\gamma^{\mu}\gamma^{\nu} -
\gamma^{\nu}\gamma^{\mu})/2$.

The effect of the universal minimal coupling between matter of any
kind and the electromagnetic field using the metric $\hat
e_{\mu\nu}$ provides an effective magnetic moment that depends on
the mass of the spinor field and on the parameter $\beta$, that is
\begin{equation}
\mu^{G} \doteq \frac{m \, c^{2}}{\beta}.
\label{1112011}
\end{equation}

Remarkably, the hypothesis of universality of $ \hat e_{\mu\nu} $
reveals the existence of an alternative origin for the magnetic
moment of all particles (charged or not)\footnote{If we had applied
the minimal coupling principle to the MM representation, we have
never obtained such term. In this representation, it corresponds to
a non-minimal coupling, which has several difficulties, as pointed
in the literature.} from first principles. The compatibility of this
result induced by the $\hat E$-representation in the
MM-representation implies that the Lagrangian describing the
interaction between the neutrino and the electromagnetic field
contains an extra term in the MM-representation which is usually
introduced by hand and without further justification. In other
words, the presence of a magnetic moment for the neutrino in the
standard MM-representation should not be viewed as an exotic
surprise but instead should be understood---on the light of the
Dynamical Bridge method---as a consequence of the universality of
the geometry $\hat e_{\mu\nu}$ in the $\hat E$-representation. Next
section, we shall introduce this map for charged particles and then
compare with experiments.

\section{The $\hat{E}-$metrical magnetic moment for charged particles}

It is important to emphasize that the value of the magnetic moment
obtained in the previous section contains only the general
contribution for any particle. Charged particles have an extra
source for $ \mu $ that is related to their charge, of course. For
instance, the standard magnetic moment of the electron is the Bohr
magneton $\mu_{B} = e \, \hbar /2 m_{e}$. Thus, the total value of
the electron magnetic moment ($\mu_e$) should be read as

$$\mu_{e} = \mu_{B} + \frac{m_{e} \, c^{2}}{\beta}+\mbox{quantum corrections}.$$
The first part corresponds to the standard magnetic moment, the
second one corresponds to our proposal, the $\hat{E}-$metrical
contribution and the last one comes from loop-quantum corrections.

\section{Comparison with experiments}
In this section we compare the contribution of the electromagnetic
metric to the anomalous magnetic moment of the leptons (electron,
muon, tau and neutrino) with observational data and phenomenology.
Actually, there are only two precise measurements for the anomaly of
the magnetic moment. One of them is the case of the electron (the
most precise measurement) which allows tests of the
quantum-electrodynamics (QED), in particular, the fine-structure
constant $\alpha$. The other one is the muon magnetic moment which
has less precise measurements, but it can test the entire standard
model of particle physics \cite{passera}. Nowadays, the anomalous
magnetic moment of the tau is unobservable due to its very short
mean life. However, there is an estimative for the value of the
anomaly that yields an upper limit of less than
$1.3\times10^{-2}\,\mu_{\tau}$. According to the standard model, the
theoretical prediction of this anomaly is
$117\,721\times10^{-8}\,\mu_{\tau}$. We will not analyze here the
case of composite particles, like protons and neutrons, since the
theory of their components (quarks and gluons) is still under
construction concerning this point \cite{roberts}.

Currently, the discrepancy between the experimental and theoretical values of the anomalous magnetic moment of the leptons is very tiny\footnote{All the values used here were taken from the \textit{Particle Data Group} \cite{pdg}}. The loop-quantum corrections of the standard magnetic moment make this discrepancy very small, restricting the possibility of new physics. The value of the so-called anomalous magnetic moment $a_l$ of the particle $l=(e,\mu,\tau)$ is defined by

\begin{equation}
\label{ano_exp}
a_l\equiv\fracc{g_l-2}{2}=\frac{m_l}{m_e} \, \frac{\mu_l}{\mu_B} -1,
\end{equation}
where $\mu_B$ is the Bohr magneton, $m_{e}$ is the electron mass,
$g_l$ is the Land\'e factor and $m_l$ is the mass of the particle
under consideration.

For muons \cite{marciano}, the difference $\Delta$ between the
experimental $E$ and the theoretical $T$ values of the anomaly is

\begin{equation}
\label{dif_muon}
\Delta a_{\mu}\equiv a^E_{\mu}-a^T_{\mu}=2.87(63)(49) \times 10^{-9}.
\end{equation}
Therefore, this is the upper bound in which the geometrical magnetic
moment $\mu^G$ may contribute. So, if one considers that the effect
of the matter coupling with $\hat e_{\mu\nu}$ appears at this order
of magnitude for this particle, one gets

\begin{equation}
\label{mu_diff_muon}
\mu^{G}_{\mu}\equiv \Delta a_{\mu} \, \fracc{e\hbar}{2m_{\mu}}.
\end{equation}

On the other hand, the geometrical magnetic moment for muons is given by the formula

\begin{equation}
\label{mu_q_muon}
\mu^{G}_{\mu}=\fracc{m_{\mu}c^2}{\beta}.
\end{equation}
If one assumes that the geometrical magnetic moment provides all the
remaining terms for the muon anomaly, it is possible to estimate
from our formula (\ref{mu_q_muon}) the critical electromagnetic
field, that is
$$\beta\approx 1.31\times10^{23}T.$$
Assuming that the calibration of the critical field given by this
equation is always valid, the geometrical corrections to the
anomalous magnetic moment of the electron is forecasted to be

\begin{equation}
\label{dif_q_elec}
\mu^{G}_{e}= 6.74\times10^{-14}\mu_B.
\end{equation}
The difference between experimental and theoretical values in this
case is about 600 times smaller than in the previous one
\cite{gabrielse}, namely
\begin{equation}
\label{dif_elec}
\Delta a_e\equiv a^E_e-a^T_e=-(0.40\pm0.88)\times10^{-12}.
\end{equation}

As one can see, the experimental errors are compatible with the
effects induced by the electromagnetic geometry upon the electron
anomaly. Concerning the tau anomaly, it is predicted to be less than
$1.3 \times 10^{-2}$, which is also in accordance with our
predictions. Indeed, using the formula (\ref{1112011}), we have
\begin{equation}
\label{dif_tau}
\mu^{G}_{\tau}= 8.15\times10^{-7}\mu_{\tau}.
\end{equation}
Despite of none precision experiments in this case, it should be
remarked that this value provides a very important prediction of our
proposal, because if the tau anomaly appears in a higher order than
the quantum corrections, it suggests a confirmation of the
geometrical magnetic moment. Otherwise, our model could be rejected
by experiments.

Let us now discuss the exceptional case of neutrinos. As uncharged
elementary particle, we could expect that neutrinos do not have any
magnetic moment. Notwithstanding, the one-loop-correction of the
standard model predicts a nonzero magnetic moment for this particle,
which depends linearly on its mass.

From the discussion above, it seems very natural to propose that the
existence of the neutrino magnetic moment has nothing to do with
quantum corrections, but instead has only the geometrical origin.
Then, we can estimate the value of the neutrino magnetic moment,
which is
\begin{equation}
\label{dif_nu}
\mu^G_{\nu}=\fracc{m_{\nu}c^2}{\beta}=1.32\times10^{-19}\mu_B.
\end{equation}
This value is compatible with the range of values obtained from
distinct areas of physics
$4\times10^{-20}\mu_B<\mu_{\nu}<9\times10^{-9}\mu_B$
\cite{balant,lych,alex}. Finally, remark that our prediction is very
close to the one established by the minimal extended standard model
($\mu_{\nu}\approx 3.2\times10^{-19}\mu_B$).

\section{Conclusions}
The possibility of describing the dynamics of a nonlinear
Born-Infeld theory in terms of the linear Maxwell theory is
certainly an unexpected result. As demonstrated along the text, the
price to pay is to introduce a curved geometry which has only an
electromagnetic character. In the present work, we tried to go ahead
by generalizing this equivalence between dynamics in the presence of
matter. This is achieved by the application of the minimal coupling
principle of any kind of matter (charged or not) with the geometry
$\hat e_{\mu\nu}.$ This implies several consequences that we have
explored only in the weak field regime, i.e., very far from the
critical field $\beta$. The most important outcome is the
contribution to the anomalous magnetic moment of leptons and,
particularly, for neutrinos. The comparison of our approach to the
experimental data indicates that our theory is not in contradiction
with observations. This model should thus be further investigated
and the case of high electromagnetic fields should be examined in
more details. We will come back to these questions elsewhere.

\subsection*{Acknowledgments}
We would like to acknowledge J.A. Helay\"el-Neto, N. Pinto-Neto,
F.T. Falciano, E. Goulart and G.B. Santos for many interesting
conversations and the staff of ICRANet in Pescara where this work
was partially done. We would like to thank FINEP, FAPERJ, CNPq and
CFC/CBPF for their financial support.
\appendix

\section{Some useful formulas on the road to the Dynamical Bridge}\label{apb}
 The two scalar invariants of the electromagnetic field $F_{\mu\nu}$
are defined as
\begin{eqnarray*}
F&\equiv&F^{\mu \nu}F_{\mu \nu} \\
G&\equiv&{}^{*}F^{\mu \nu}F_{\mu \nu}.
\end{eqnarray*}
From these definitions the following algebraic relations follows:

\begin{eqnarray}
&&{}^{*}F^{\mu \alpha}{}^{*}F_{\alpha \nu}-F^{\mu \alpha}F_{\alpha \nu}=\frac{F}{2}\delta^\mu{}_{\nu},\label{algrel1}\\
&&{}^{*}F^{\mu \alpha}F_{\alpha \nu}=-\frac{G}{4}\delta^\mu{}_{\nu},\label{algrel2}\\
&&F^\mu{}_\alpha F^\alpha{}_\beta F^\beta{}_\nu=-\frac{G}{4}{}^{*}F^{\mu}{}_{\nu}-\frac{F}{2}F^\mu{}_{\nu},\label{algrel3}\\
&&F^\mu{}_\alpha F^\alpha{}_\beta F^\beta{}_\lambda F^\lambda{}_\nu=\frac{G^2}{16}\delta^\mu{}_{\nu}-\frac{F}{2}F^\mu{}_{\alpha}F^\alpha{}_{\nu}. \quad \;\label{algrel4}
\end{eqnarray}

In the Maxwell theory, every tensor index is raised and lowered
using the Minkowski metric $\eta_{\mu \nu}$, while in the curved
$\hat{E}-$ metric the indexes must be raised and lowered using $\hat
e_{\mu \nu}$. In a schematic way, we have:

\begin{itemize}
\item{\textbf{Maxwell's theory}: it is represented in the Minkowski metric $ \eta_{\mu \nu}$ and the Lagrangian is $L=-F/4$. Then,
\begin{eqnarray*}
&&F^{\mu \nu} \equiv F_{\alpha \beta}\eta^{\mu \alpha}\eta^{\nu \beta},\\[2ex]
&&F = F_{\mu \nu}F_{\alpha \beta}\eta^{\mu \alpha}\eta^{\nu \beta}.
\end{eqnarray*}
The dynamical equation of this theory in the absence of sources is
given by
$$\fracc{1}{\sqrt{-\eta}}(\sqrt{-\eta}F_{\alpha \beta}\eta^{\mu \alpha}\eta^{\nu \beta})_{,\nu}=0.$$
}
\item{\textbf{Born-Infeld's theory}: it corresponds to the curved metric $\hat e_{\mu \nu}$ and the nonlinear Lagrangian $\hat{L}=\beta^2\left(1-\sqrt{\hat{U}} \right)$, where $\beta$ is a free parameter (critical electromagnetic field) and
$$\hat{U}=1+\frac{\hat{F}}{2\beta^2}-\frac{\hat{G}}{16\beta^4}.$$
We define
\begin{eqnarray*}
&&\hat{F}^{\mu\nu}\equiv F_{\alpha \beta}\hat{e}^{\mu \alpha}\hat{e}^{\nu \beta},\\[2ex]
&&\hat{F}=F_{\mu \nu}F_{\alpha \beta}\hat{e}^{\mu \alpha}\hat{e}^{\nu \beta},\\[2ex]
&&\hat{G}=\frac{1}{\sqrt{-\hat{e}}}\,\epsilon^{\mu \nu \alpha \beta} F_{\mu \nu}F_{\alpha \beta}={}^{*}\hat F{}^{\mu\nu}F_{\mu \nu},
\end{eqnarray*}
where $\epsilon^{\mu \nu \alpha \beta}$ is the completely
skew-symmetric object and $\epsilon^{0123}=1$. The dynamics of this
nonlinear theory in the absence of source is provided by
$$\partial_{\nu}\left[ \frac{\sqrt{- \, \hat e }}{\hat U } \left(  \hat F^{\mu\nu} - \frac{1}{4 \, \beta^{2}} \, \hat G \, ^{*}\hat F^{\mu\nu}\right)  \right] = 0.$$
}
\end{itemize}

The electromagnetic metric $\hat e_{\mu \nu}$ depends on the
electromagnetic field and the background geometry. Due to the
algebraic relations\ (\ref{algrel1})-(\ref{algrel4}), there is a
unique way to define the electromagnetic metric, namely,
\begin{equation}
\label{qmet}
\hat e^{\mu \nu}=a \, \eta^{\mu \nu}+b \, \Phi^{\mu \nu},
\end{equation}
where $a$ and $b$ are functions of the Lorentz invariants $F$ and
$G$ and $\Phi^{\mu \nu}\equiv F^{\mu \alpha}F_{\alpha}{}^{\nu}$. The
term electromagnetic metric is due to the fact that $\hat e_{\mu
\nu}$ is defined uniquely in terms of the electromagnetic fields.
 The relation between the scalar invariants of these theories is
obtained according to:

\begin{equation}
\label{inv_rel}
\left\{\begin{array}{lcl}
\hat F&=&a^2\left[(p-\epsilon Fq)F-\fracc{\epsilon q}{2}G^2\right],\\[2ex]
\hat G&=&a^2\left[p-\fracc{\epsilon q}{2}F\right]G,
\end{array}\right.
\end{equation}
where
$$p\equiv 1+\fracc{\epsilon^2G^2}{16},\hspace{1cm}q\equiv1-\fracc{\epsilon F}{4}.$$
 These are the necessary formula to prove the dynamical bridge.

\section{General expression for the spin connection}\label{apa}

From the condition\ (\ref{1235}) the internal connection
$\hat\Gamma_{\mu}$ is provided by

\begin{equation}
\label{int_conn}
\hat\Gamma_{\mu}=\hat\Gamma^{FI}_{\mu}+\hat V_{\mu},
\end{equation}
where $\hat V_{\mu}$ is an arbitrary vector satisfying the Clifford
algebra and $\hat\Gamma^{FI}_{\mu}$ is the Fock-Ivanenko connection
defined in terms of the Dirac matrices $\hat\gamma^{\mu}$'s, as
follows

\begin{equation}
\label{fi_conn}
\hat\Gamma^{FI}_{\mu} \equiv \fracc{1}{8} \left[ \hat\gamma^{\lambda} \hat\gamma_{\lambda,\mu} - \hat\gamma_{\lambda,\mu} \hat\gamma^{\lambda} + \hat\Gamma^{\epsilon}_{\lambda\mu}( \hat\gamma_{\epsilon}\hat\gamma^{\lambda}-\hat\gamma^{\lambda}\hat\gamma_{\epsilon})\right],
\end{equation}
where $\hat\Gamma^{\epsilon}_{\lambda\mu}$ is the Christoffel symbol
constructed with the $\hat e_{\mu\nu}$ metric. The covariant
derivative of $\Psi$ is given by

\begin{equation}
\label{cov_der_spin}
\begin{array}{ll}
\hat\nabla_{\mu}\Psi&=\partial_{\mu}\Psi-\hat\Gamma_{\mu}\Psi\\[2ex]
&=\partial_{\mu}\Psi-\hat\Gamma^{FI}_{\mu}\Psi-\hat V_{\mu}\Psi.
\end{array}
\end{equation}
In order to preserve the probability current $\hat J^{\mu} \equiv \bar\Psi \hat\gamma^{\mu} \Psi$ in the metric $\hat e_{\mu\nu}$, i.e,
$$\hat\nabla_{\mu}\hat J^{\mu}=0,$$
the hermitian conjugate expression of\ (\ref{cov_der_spin}) takes
the form

\begin{equation}
\label{cov_der_spin_adj}
\begin{array}{ll}
\hat\nabla_{\mu}\bar\Psi&=\partial_{\mu}\bar\Psi+\bar\Psi\hat\Gamma_{\mu}\\[2ex]
&=\partial_{\mu}\bar\Psi+\bar\Psi\hat\Gamma^{FI}_{\mu}+\bar\Psi \hat V_{\mu}.
\end{array}
\end{equation}
In the general case, it was demonstrated in \cite{formiga} that
$\hat V^{\mu}$ must transform like the Fock-Ivanenko connection in
order to verify the conservation laws. Thus, according to the choice
of $\hat V^{\mu}$ given by\ (\ref{choice_v}), that contains the
anti-symmetric tensor $F_{\mu\nu}$, guarantees the conservation law.
From these statements, it is straightforward to verify that the
probability current is also conserved in Minkowski space.


\begin{thebibliography}{45}
\bibitem{sanda}
W.J. Marciano and A.I. Sanda, {\em Phys.\ Lett.\ B} {\bf 67} 303 (1977).
\bibitem{lee}
B.W. Lee and R.E. Shrock, {\em Phys.\ Rev.\ D} {\bf 16} 1444 (1977).
\bibitem{helayel}
H. Belich, L.P. Colatto, T. Costa-Soares, J.A. Helay\"el-Neto and M.T.D. Orlando, {\em Eur.\ Phys.\ J.\ C} {\bf 62} 425 (2009).
\bibitem{kau}
K. Bhattacharya and P.B. Pal, {\em Proc.\ Indian\ Natn\ Sci\ Acad} {\bf 70, A} 145 (2004).
\bibitem{novello et al}
M. Novello, F.T. Falciano and E. Goulart, arXiv:gr-qc/1111.2631 (2011).
\bibitem{born}
M. Born, {\em Nature (London)} {\bf 132} 282 (1933); {\em Proc.\ R.\ Soc.\ A} {\bf 143} 410 (1934); M. Born and L. Infeld, {\em Proc.\ R.\ Soc.\ A} {bf 144} 425 (1934).
\bibitem{novello_73}
M. Novello, {\em Phys.\ Rev.\ D}, {\bf 8} 8, 2398 (1973).
\bibitem{shapos}
M. Shaposhnikov, {\em Proceeding of the 11th Marcel Grossmann Meeting on General Relativity }, p. 1006, Berlin, Germany (2006).
\bibitem{geng}
C.Q. Geng and R. Takahashi, {\em Phys.\ Lett.\ B} {\bf 710} 324 (2012).
\bibitem{pdg}
J. Beringer et al., \textit{Particle Data Group}, {\em Phys.\ Rev.\ D} {\bf 86} 010001 (2012).
\bibitem{passera}
M. Passera, {\em Nucl.\ Phys.\ Proc.\ Suppl.} {\bf 169} 213 (2007).
\bibitem{roberts}
I.C. Cloet, C.D. Roberts and A.W. Thomas, arXiv:nucl-th/1304.0855 (2013).
\bibitem{marciano}
A. Czarneck and W.J. Marciano, {\em Phys.\ Rev.\ D} {\bf 64} 013014 (2001).
\bibitem{gabrielse}
D. Hanneke, S. Fogwell, and G. Gabrielse, {\em Phys.\ Rev.\ Lett.} {\bf 100} 120801 (2008).
\bibitem{balant}
A.B. Balantekin, {\em Proceedings of Origin of Matter and the Evolution of Galaxies (OMEG05)}, Tokyo, Japan (2005).
\bibitem{lych}
O. Lychkovskiy and S. Blinnikov, {\em Phys.\ Atom.\ Nucl.} {\bf 73} 614 (2010).
\bibitem{alex}
A.V. Kuznetsov and N.V. Mikheev {\em JCAP} {\bf 11} 031 (2007).
\bibitem{formiga}
J.B. Formiga, arXiv:hep-th/1210.0759.
\end{thebibliography}
\end{document}